\documentclass[oldversion]{aa}

\usepackage[round]{natbib}
\usepackage{times,float}
\usepackage[T1]{fontenc}
\usepackage{epsfig}
\usepackage{graphics}
\usepackage{amssymb}
\usepackage{txfonts}

\begin{document}

\title{BOOTES Observation of GRB\,080603B}

\author{Martin Jel\'{\i}nek\inst{1}
\and Javier Gorosabel\inst{1}
\and Alberto J. Castro-Tirado\inst{1}
\and Antonio de Ugarte Postigo\inst{1} \and \\
 Sergei Guziy\inst{1}
\and Ronan Cunniffe\inst{1}
\and Petr Kub\'anek\inst{2}
\and Michael Prouza\inst{2} \and 
 Stanislav V\'{\i}tek\inst{3} \and \\
 Ren\'e Hudec\inst{4,3}
\and Victor Reglero\inst{5}
\and Lola Sabau-Graziati\inst{6}
}
\institute{Instituto de Astrof\'{\i}sica de Andaluc\'{\i}a CSIC, Granada, Spain \and
Fyzik\'aln\'{\i} \'ustav (Fz\'U AV \v{C}R), Praha, Czech Republic
\and Fakulta Elektrotechnick\'a, \v CVUT v Praze, Czech Republic
\and Astronomick\'y \'ustav Akademie v\v{e}d (AS\'U AV \v CR), Ond\v{r}ejov, Czech Republic
\and Image Processing Laboratory, Universitat de Valencia, Spain
\and Instituto Nacional de T\'ecnica Aeroespacial, Torrej\'on de Ardoz, Madrid, Spain
}



\abstract{We report on multicolor photometry of long GRB\,080603B afterglow from
BOOTES-1B and BOO\-TES-2. The optical afterglow has already been reported to
present a break in the optical lightcurve at 0.12$\pm$0.2 days after the
trigger. We construct the lightcurve and the spectral energy distribution and
discuss the nature of the afterglow.}


\offprints{ \hbox{Martin Jel\'{\i}nek, e-mail:\,{\tt mates@iaa.es}}}

\maketitle

\date{Received  / Accepted }

\section{Introduction}

GRB\,080603B was a long gamma-ray burst detected on June 3, 2008, at
19:38:13\,UT by {\it Swift}-BAT \citep{gcnreport}. The burst was also detected
by {\it Konus}-WIND \citep{konus} and {\it INTEGRAL}-SPI/ACS
\citep{spiacs}. 

In X-rays, the afterglow was detected by {\it Swift}-XRT, providing a rapid and
precise localization \citep{xrt}. 

The optical afterglow was observed by several telescopes - ROTSE III
\citep{rotse}, TAROT (\cite{tarot}, \cite{tarot2}, \cite{tarot3}), TLS
Tautenburg \citep{tautenburg}, RTT150 \citep{turkish}, the Liverpool Telescope
\citep{lt}, Xinglong EST \citep{xing}, the 1.0m telescope at CrAO \citep{crao,
crao2}, the 1.5\,m telescope of Sayan observatory \citep{sayan} and from Maidanak \citep{mao}. In infrared by PAIRITEL \citep{pairitel}, spectroscopy was
obtained by the NOT \citep{not} and the Hobby-Eberly Telescope \citep{hobby},
providing a redshift of z=2.69. 

An upper limit on radio emission was set by the VLA \citep{vla}. 

\section{Observations}

At both BOOTES stations, the GRB happened during twilight, delaying
follow-up by $\sim$ 1h. Despite the delay, the optical afterglow is well
detected in the data from both telescopes.

The 60\,cm telescope BOOTES-2/TELMA, in La Mayora, M\'alaga, Spain, started
taking data at 20:29:19\,UT, i.e.  51 minutes after the GRB trigger. A sequence
of r$'$-band exposures was taken, and later, after confirming the detection of
the optical transient, i$'$, g$'$ and Y band images were obtained. In the near
infrared Y band, despite 600\,s of integration, the afterglow was not detected. 

The 30\,cm telescope BOOTES-1B, located in El Arenosillo, Huelva, Spain,
\citep{boobible} obtained 368 unfiltered images totalling more than 6 hours of
integrated light until the end of the night. The images were combined to
improve signal-to-noise, to yield 11 data points for the period between 1.2 and
5.2 hours after the GRB. One point has a large error due to clouds crossing the
field of view.

Best fit astrometric position of the afterglow, obtained from the weighted
average of all available images from BOOTES-2 is 
$$ \alpha = 11:46:07.73 \quad \delta = +68:03:39.9 \quad (J2000),$$
	about 1.6$^\circ$ SE from star $\lambda$\,Dra. 

\begin{figure} \begin{center} \label{img} \resizebox{0.707\hsize}{!}{
\includegraphics{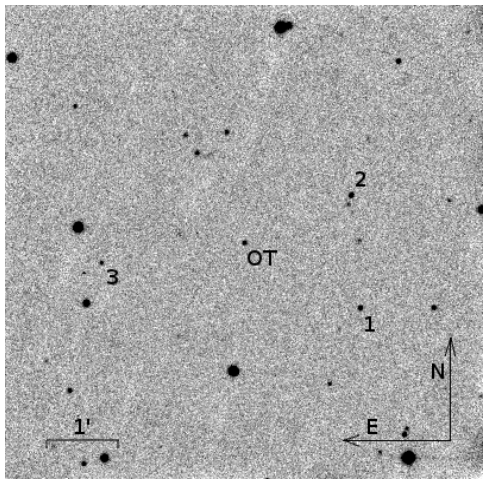} }
	\caption{The finding chart of the afterglow of GRB080603B. Combination
of images taken by BOOTES-2.}
	\end{center} 

\end{figure}

Photometry was done in the optimal aperture using IRAF/Daophot. Calibration was
performed against three SDSS (DR8)\citep{sdss-dr8} stars. The stars are marked
on the identification chart (Fig.\,\ref{img}) and their brightnesses are in the
Table\,\ref{table-calib}. Our unfiltered, "Clear", best fit magnitude
Clear=$A_1*g'+A_2*r'$ used for BOOTES-1B calibration is mentioned as well. 

For the summary of our observations, see Table\,\ref{table-data}.

\begin{table}[b!]
\begin{center} 
\caption{ Calibration stars used.}
\label{table-calib}

\begin{tabular*}{0.4\hsize}{@{\extracolsep{\fill}}ccccc}
\\
\hline 
{\normalsize ID.} & 
{\normalsize g$'$} & 
{\normalsize r$'$} & 
{\normalsize i$'$} & 
{\normalsize Clear} 
\\ \hline 

1 & 18.00 & 17.50 & 17.32 & 17.52 \\
2 & 18.80 & 17.35 & 16.04 & 17.35 \\
3 & 19.88 & 18.42 & 17.09 & 18.47 \\

\hline

\end{tabular*}
\end{center} 
\end{table}

\begin{table*}
\begin{center} 
\caption{Optical photometric observations of the
optical afterglow of the GRB\,080603B.} 
\label{table-data}

\begin{tabular*}{0.841\hsize}{@{\extracolsep{\fill}}ccccccc}
\\
\hline 
{\normalsize UT Date of mid exp.} & 
{\normalsize $T- T_0 [h]$} & 
{\normalsize tel.} &
{\normalsize filter} &
{\normalsize $T_{\rm exp}[s]$} &
{\normalsize mag} & 
{\normalsize $\delta$ mag} 
\\ \hline 

Jun 3.855805 & 0.902 & B-2  & r$'$  & 3$\times$120\,s & 17.46 & 0.07 \\
Jun 3.859348 & 0.987 & B-2  & r$'$  & 2$\times$120\,s & 17.59 & 0.13 \\
Jun 3.862188 & 1.056 & B-2  & r$'$  & 2$\times$120\,s & 17.31 & 0.05 \\
Jun 3.864311 & 1.107 & B-2  & r$'$  & 120\,s          & 17.57 & 0.08 \\
Jun 3.865747 & 1.141 & B-2  & r$'$  & 120\,s          & 17.30 & 0.07 \\
Jun 3.867151 & 1.175 & B-2  & r$'$  & 120\,s          & 17.46 & 0.06 \\
Jun 3.868946 & 1.218 & B-1B & Clear & 10$\times$60\,s & 17.53 & 0.07 \\
Jun 3.870011 & 1.243 & B-2  & g$'$  & 3$\times$120\,s & 18.29 & 0.04 \\
Jun 3.874248 & 1.345 & B-2  & g$'$  & 3$\times$120\,s & 18.24 & 0.04 \\
Jun 3.876758 & 1.405 & B-1B & Clear & 10$\times$60\,s & 17.54 & 0.06 \\
Jun 3.879225 & 1.465 & B-2  & g$'$  & 4$\times$120\,s & 18.14 & 0.03 \\
Jun 3.884248 & 1.585 & B-2  & r$'$  & 3$\times$120\,s & 17.50 & 0.09 \\
Jun 3.884664 & 1.595 & B-1B & Clear & 10$\times$60\,s & 17.70 & 0.06 \\
Jun 3.889912 & 1.721 & B-2  & r$'$  & 3$\times$120\,s & 17.70 & 0.15 \\
Jun 3.892654 & 1.787 & B-1B & Clear & 10$\times$60\,s & 17.75 & 0.06 \\
Jun 3.893455 & 1.806 & B-2  & r$'$  & 4$\times$120\,s & 17.74 & 0.06 \\
Jun 3.899839 & 1.959 & B-2  & g$'$  & 5$\times$120\,s & 18.42 & 0.19 \\
Jun 3.900620 & 1.978 & B-1B & Clear & 10$\times$60\,s & 17.79 & 0.06 \\
Jun 3.906961 & 2.130 & B-2  & g$'$  & 5$\times$120\,s & 18.42 & 0.04 \\
Jun 3.908509 & 2.167 & B-1B & Clear & 10$\times$60\,s & 17.87 & 0.09 \\
Jun 3.914867 & 2.320 & B-2  & r$'$  & 4$\times$120\,s & 18.15 & 0.13 \\
Jun 3.916482 & 2.359 & B-1B & Clear & 10$\times$60\,s & 17.91 & 0.11 \\
Jun 3.922694 & 2.508 & B-2  & i$'$  & 5$\times$120\,s & 17.89 & 0.05 \\
Jun 3.931774 & 2.726 & B-2  & r$'$  & 7$\times$120\,s & 18.01 & 0.06 \\
Jun 3.934988 & 2.803 & B-1B & Clear & 35$\times$60\,s & 18.30 & 0.32 \\
Jun 3.940845 & 2.943 & B-2  & i$'$  & 5$\times$120\,s & 17.88 & 0.07 \\
Jun 3.947882 & 3.112 & B-2  & r$'$  & 5$\times$120\,s & 18.12 & 0.08 \\
Jun 3.956941 & 3.330 & B-1B & Clear & 20$\times$60\,s & 18.45 & 0.07 \\
Jun 3.971736 & 3.685 & B-1B & Clear & 21$\times$60\,s & 18.38 & 0.06 \\
Jun 3.977109 & 3.814 & B-2  & r$'$  & 5$\times$120\,s & 18.26 & 0.18 \\
Jun 4.006997 & 4.531 & B-1B & Clear & 78$\times$60\,s & 18.79 & 0.07 \\

\hline

\end{tabular*}
\end{center} 
\end{table*}

\begin{figure} \begin{center} \label{lc} \resizebox{\hsize}{!}{
\includegraphics{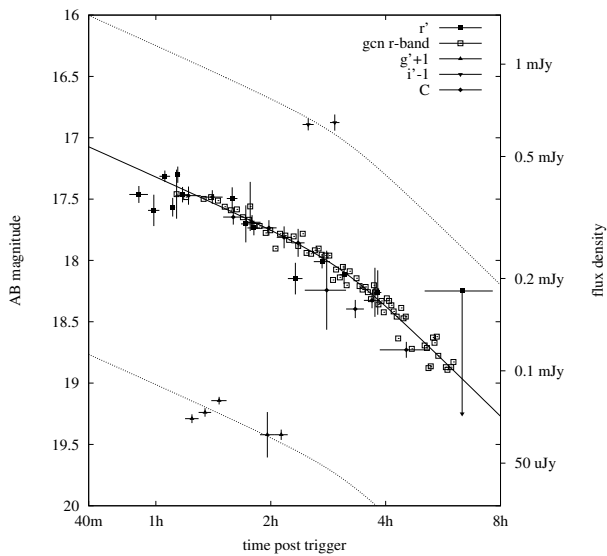} }
	\caption{The detail of the optical light curve of GRB\,080603B showing
the observations by BOOTES (filled symbols) and from literature (empty
symbols).}
	\end{center} 

\end{figure}

\begin{figure}
        \begin{center} \label{lc2}
        \resizebox{\hsize}{!}{
\includegraphics{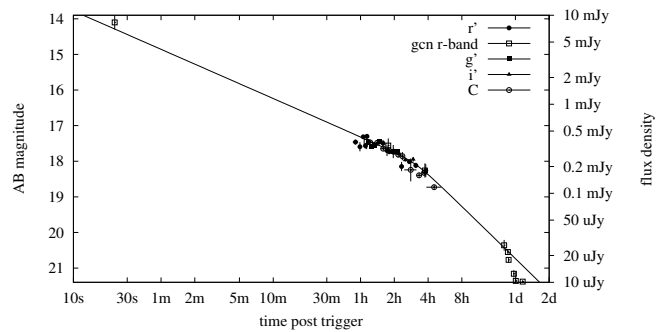}
	}
        \caption{The overall view of the light curve of GRB\,080603B.}
        \end{center} 

\end{figure}

\begin{figure} \begin{center} \label{sed} \resizebox{\hsize}{!}{
\includegraphics{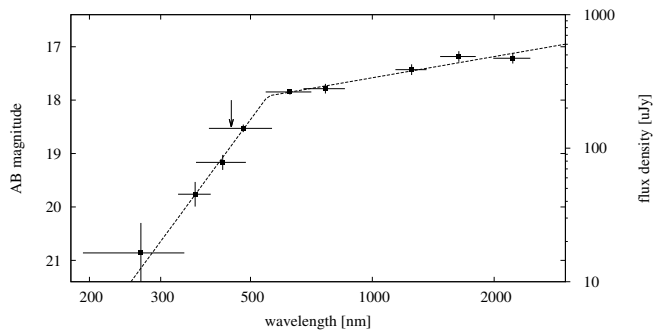} }
	\caption{The spectral energy distribution of the afterglow in rest
frame. The small arrow marks Ly-$\alpha$ position for $z=2.69$.  }
	\end{center} 

\end{figure}

\section{Fitting The Lightcurve}

The lightcurve, as already shown by \cite{turkish} shows a smooth transition
between two decay slopes $\alpha_1=-0.55\pm0.16$ and $\alpha_2=-1.23\pm0.22$.
The break occurs at $t_b = 0.129\pm0.016$\,days. 

There is no hint of chromatic evolution within the lightcurve so all filters
were scaled and fitted together with the $r'$-band. 
The fitting of the lightcurve was performed in $\log t / \log f$ space, where
power law functions, typical for gamma-ray bursts, show as straight lines. We
used a hyperbolic transition between two slopes (smoothly broken power-law):
$$ h(a,b)=a + \frac{b}{2} \sqrt{1+\frac{a^2}{b^2}} $$
$$ m(t) = m_0 - 2.5 \alpha_2 \log \frac{t}{ t_b} \ + h( - 2.5
(\alpha_1-\alpha_2)\log \frac{t}{t_b} , G ) $$
Where $\alpha_1$ and $\alpha_2$ are pre-break and post-break decay indices,
$t_b$ is the break time, $m_0$ is an absolute scaling parameter of the
brightness and $G$ expresses smoothness of the break. 

Although the early point by ROTSE \citep{rotse} was not used, it agrees with the
backward extrapolation of the $\alpha_1$ slope and so supports this simple
interpretation. 

We constructed a spectral energy distribution (SED) by fitting the needed
magnitude shift of the R-band lightcurve model to the photometric points from
BOOTES, UVOT \citep{gcnreport} and PAIRITEL \citep{pairitel} obtained in other
filters. While the points from UVOT are practically contemporaneous to BOOTES,
PAIRITEL observed rather later (0.32\,days after trigger), so the SED is
therefore model-dependent in its infrared part. The synthetic AB magnitudes
equivalent to $t=0.1$\,days are in Table\,\ref{table-sed}.

\begin{table}[h!]
\begin{center} 
\caption{The spectral energy distribution in AB magnitudes equivalent to
$0.1$\,days after the trigger. ($^\dagger$ UVOT, $^\ddagger$ PAIRITEL)}
\label{table-sed}

\begin{tabular*}{0.4\hsize}{@{\extracolsep{\fill}}ccc}
\\
\hline 
{\normalsize Filter} & 
{\normalsize $m_{\mathrm{AB}}$} & 
{\normalsize $\Delta m_{\mathrm{AB}}$} 
\\ \hline 

W$^\dagger$  & 20.98 & 0.56 \\
U$^\dagger$  & 19.83 & 0.23 \\
B$^\dagger$  & 19.22 & 0.14 \\
g'          & 18.57 & 0.07 \\
r'          & 17.88 & 0.05 \\
i'          & 17.81 & 0.09 \\
J$^\ddagger$ & 17.44 & 0.10 \\
H$^\ddagger$ & 17.19 & 0.10 \\
K$^\ddagger$ & 17.22 & 0.10 \\

\hline

\end{tabular*}
\end{center} 
\end{table}

The SED shows a clear supression of radiation above ~4500$\AA$, i.e. redshifted
Ly-$\alpha$ line. 
No radiation is detected above the Lyman break at 3365$\AA$.
A rather shallow power law with an index $\beta = -0.53 \pm 0.06$ was found
redwards from $r'$ band. The fit was performed using the E(B-V) = 0.013\,mag
\citep{dust}.

The strong suppression of light for wavelengths shorter than r' band is likely
due to the Ly-$\alpha$ absorption within the host galaxy and Ly-alpha line
blanketing for $z$=2.69.

\section{Discussion}

The values of $\alpha_2=-1.23\pm0.22$ and $\beta = -0.53 \pm 0.06$ both point to
a common electron distribution parameter $p=2.05\pm0.20$ ($\alpha=(3*p-1)/4$,
$\beta=(p-1)/2$) \citep{piran-grb-physics}. Such a combination suggests a
stellar wind profile expansion and a slow cooling regime. 

The pre-break decay rate $\alpha_1=-0.55\pm0.16$ remains unexplained by the
standard fireball model. It is unlikely that the break at $t_b = 0.129\pm0.016$
would be a jet break. It is quite possible that the plateau is not really a
straight power law, and that some late activity of the inner engine may be
producing bumping of hydrodynamic origin.

We note that the literature contains a number of observations suggesting a
rapid decay by about one day after the GRB. Without having all the images, it
is, however, impossible to decide whether this is a real physical effect or a
zero-point mismatch. 

\section{Conclusions}

The 0.6\,m telescope BOOTES-2 in La Mayora observed the optical afterglow of
GRB\,080603B in three filters. The 0.3\,m BOOTES-1B in El Arenosillo observed
the same optical afterglow without filter.

Using the data we obtained at BOOTES and from the literature, we construct the
lightcurve and broadband spectral energy distribution. 

Our fit of the obained data privides the decay parameters
$\alpha_2=1.23\pm0.22$ and $\beta = -0.53 \pm 0.06$, which suggest a slow
cooling expansion into a stellar wind. 

\begin{acknowledgement}We acknowledge the support of the Spanish Ministerio de
Ciencia y Tecnolog\'\i a through Projects AYA2008-03467/ESP and
AYA2009-14000-C03-01/ESP, and Junta de Andaluc\'\i a through the Excellence
Reseach Project P06-FQM-219, and the GA\v{C}R grants 205/08/1207 and
102/09/0997. We are also indebted to T. Mateo-Sanguino (UHU), J.A. Adame, J. A.
Andreu, B. de la Morena, J. Torres (INTA) and to R.  Fern\'andez-Mu\-noz
(EELM-CSIC), V. Mu\-noz-Fern\'andez and C. P\'erez del Pulgar (UMA) for their
support.  \end{acknowledgement}

\bibliography{mj080603B}

\hyphenation{Post-Script Sprin-ger}
\begin{thebibliography}{24}
\providecommand{\natexlab}[1]{#1}
\providecommand{\url}[1]{\texttt{#1}}
\expandafter\ifx\csname urlstyle\endcsname\relax
  \providecommand{\doi}[1]{doi: #1}\else
  \providecommand{\doi}{doi: \begingroup \urlstyle{rm}\Url}\fi

\bibitem[{Chandra} and {Frail}(2008)]{vla}
P.~{Chandra} and D.~A. {Frail}.
\newblock {GRB\,080603B: PAIRITEL infrared detection}.
\newblock \emph{GCN Circular 7827}, 2008.

\bibitem[{Cucchiara} and {Fox}(2008)]{hobby}
A.~{Cucchiara} and D.~{Fox}.
\newblock {GRB\,080603B: Hobby-Eberly Telescope redshift confirmation}.
\newblock \emph{GCN Circular 7815}, 2008.

\bibitem[{Eisenstein} et~al.(2011){Eisenstein}, {Weinberg}, {Agol}, {Aihara},
  {Allende Prieto}, {Anderson}, {Arns}, {Aubourg}, {Bailey}, {Balbinot}, and
  et~al.]{sdss-dr8}
D.~J. {Eisenstein}, D.~H. {Weinberg}, E.~{Agol}, H.~{Aihara}, C.~{Allende
  Prieto}, S.~F. {Anderson}, J.~A. {Arns}, {\'E}.~{Aubourg}, S.~{Bailey},
  E.~{Balbinot}, and et~al.
\newblock {SDSS-III: Massive Spectroscopic Surveys of the Distant Universe, the
  Milky Way, and Extra-Solar Planetary Systems}.
\newblock \emph{\aj}, 142:\penalty0 72, Sept. 2011.
\newblock \doi{10.1088/0004-6256/142/3/72}.

\bibitem[{Fynbo} et~al.(2008){Fynbo}, {Quirion}, {Xu}, {Malesani}, {Thoene},
  {Hjorth}, {Milvang-Jensen}, and {Jakobson}]{not}
J.~{Fynbo}, P.-O. {Quirion}, D.~{Xu}, D.~{Malesani}, C.~{Thoene}, J.~{Hjorth},
  B.~{Milvang-Jensen}, and P.~{Jakobson}.
\newblock {GRB\,080603B: NOT redshift}.
\newblock \emph{GCN Circular 7797}, 2008.

\bibitem[{Golenetskii} et~al.(2008){Golenetskii}, {Aptekar}, {Mazets},
  {Pal'shin}, {Frederiks}, and Cline]{konus}
S.~{Golenetskii}, R.~{Aptekar}, E.~{Mazets}, V.~{Pal'shin}, D.~{Frederiks}, and
  T.~Cline.
\newblock {Konus-Wind observation of GRB\,080603B}.
\newblock \emph{GCN Circular 7812}, 2008.

\bibitem[{Ibrahimov} et~al.(2008){Ibrahimov}, {Karimov}, {Rumyantsev}, and
  {Pozanenko}]{mao}
M.~{Ibrahimov}, P.~{Karimov}, A.~{Rumyantsev}, and A.~{Pozanenko}.
\newblock {GRB\,080603B: optical observations in MAO}.
\newblock \emph{GCN Circular 7975}, 2008.

\bibitem[{Jel{\'{\i}}nek} et~al.(2010){Jel{\'{\i}}nek}, {Castro-Tirado}, {de
  Ugarte Postigo}, {Kub{\'a}nek}, {Guziy}, {Gorosabel}, {Cunniffe},
  {V{\'{\i}}tek}, {Hudec}, {Reglero}, and {Sabau-Graziati}]{boobible}
M.~{Jel{\'{\i}}nek}, A.~J. {Castro-Tirado}, A.~{de Ugarte Postigo},
  P.~{Kub{\'a}nek}, S.~{Guziy}, J.~{Gorosabel}, R.~{Cunniffe},
  S.~{V{\'{\i}}tek}, R.~{Hudec}, V.~{Reglero}, and L.~{Sabau-Graziati}.
\newblock {Four Years of Real-Time GRB Followup by BOOTES-1B (2005-2008)}.
\newblock \emph{Advances in Astronomy}, 2010:\penalty0 432172, 2010.
\newblock \doi{10.1155/2010/432172}.

\bibitem[{Kann} et~al.(2008){Kann}, {Laux}, and {Ertel}]{tautenburg}
D.~{Kann}, U.~{Laux}, and S.~{Ertel}.
\newblock {GRB\,080603B: TLS Afterglow Observation}.
\newblock \emph{GCN Circular 7823}, 2008.

\bibitem[{Klotz} et~al.(2008{\natexlab{a}}){Klotz}, {Boer}, and
  {Atteia}]{tarot}
A.~{Klotz}, M.~{Boer}, and J.~{Atteia}.
\newblock {GRB\,080603B: TAROT Calern observatory detection of a plateau in the
  light curve}.
\newblock \emph{GCN Circular 7795}, 2008{\natexlab{a}}.

\bibitem[{Klotz} et~al.(2008{\natexlab{b}}){Klotz}, {Boer}, and
  {Atteia}]{tarot2}
A.~{Klotz}, M.~{Boer}, and J.~{Atteia}.
\newblock {GRB\,080603B: TAROT Calern observatory confirmation of slow optical
  decay}.
\newblock \emph{GCN Circular 7799}, 2008{\natexlab{b}}.

\bibitem[{Klotz} et~al.(2009){Klotz}, {Bo\"er}, {Atteia}, and {Gendre}]{tarot3}
A.~{Klotz}, M.~{Bo\"er}, J.~{Atteia}, and B.~{Gendre}.
\newblock {Early Optical Observations of Gamma-Ray Bursts by the TAROT
  Telescopes: Period 2001 -- 2008}.
\newblock \emph{The Astronomical Journal}, 2009.

\bibitem[{Klunko} and {Pozanenko}(2008)]{sayan}
E.~{Klunko} and A.~{Pozanenko}.
\newblock {GRB\,080603B: optical observation}.
\newblock \emph{GCN Circular 7890}, 2008.

\bibitem[{Mangano} et~al.(2008{\natexlab{a}}){Mangano}, {La Parola}, and
  {Sbarufatti}]{xrt}
V.~{Mangano}, B.~{La Parola}, and B.~{Sbarufatti}.
\newblock {GRB\,080603B: Swift-XRT refined analysis}.
\newblock \emph{GCN Circular 7806}, 2008{\natexlab{a}}.

\bibitem[{Mangano} et~al.(2008{\natexlab{b}}){Mangano}, {Parsons}, {Sakamoto},
  {La Parola}, {Kuin}, {Barthelmy}, {Burrows}, {Roming}, and
  {Gehrels}]{gcnreport}
V.~{Mangano}, A.~{Parsons}, T.~{Sakamoto}, V.~{La Parola}, N.~{Kuin},
  S.~{Barthelmy}, D.~{Burrows}, P.~{Roming}, and N.~{Gehrels}.
\newblock {Swift Observation of GRB\,080603B}.
\newblock \emph{GCN Report 144}, 2008{\natexlab{b}}.

\bibitem[{Melandri} et~al.(2008){Melandri}, {Gomboc}, {Guidorzi}, {Smith},
  {Steele}, {Bersier}, {Mundell}, {Carter}, {Kobayashi}, {Burgdorf}, {Bode},
  {Rol}, {O'Brien}, {Bannister}, and {Tanvir}]{lt}
A.~{Melandri}, A.~{Gomboc}, C.~{Guidorzi}, R.~{Smith}, I.~{Steele},
  D.~{Bersier}, C.~{Mundell}, D.~{Carter}, S.~{Kobayashi}, M.~{Burgdorf},
  M.~{Bode}, E.~{Rol}, P.~{O'Brien}, N.~{Bannister}, and N.~{Tanvir}.
\newblock {GRB\,080603B: Liverpool Telescope Observations}.
\newblock \emph{GCN Circular 7813}, 2008.

\bibitem[{Miller} et~al.(2008){Miller}, {Bloom}, and {Perley}]{pairitel}
A.~{Miller}, J.~{Bloom}, and D.~{Perley}.
\newblock {GRB\,080603B: PAIRITEL infrared detection}.
\newblock \emph{GCN Circular 7827}, 2008.

\bibitem[{Piran}(2004)]{piran-grb-physics}
T.~{Piran}.
\newblock {The physics of gamma-ray bursts}.
\newblock \emph{Reviews of Modern Physics}, 76:\penalty0 1143--1210, Oct. 2004.
\newblock \doi{10.1103/RevModPhys.76.1143}.

\bibitem[Rau(2012)]{spiacs}
A.~Rau.
\newblock {Catalogue of SPI-ACS Gamma-Ray Burst}.
\newblock
  \emph{{http://\-www.mpe.mpg.de/\-gamma/\-science/\-grb/\-1ACSburst.html}},
  2012.

\bibitem[{Rujopakarn} et~al.(2008){Rujopakarn}, {Guver}, and {Smith}]{rotse}
W.~{Rujopakarn}, T.~{Guver}, and D.~{Smith}.
\newblock {GRB\,080603B: ROTSE-III Detection of Optical Counterpart}.
\newblock \emph{GCN Circular 7792}, 2008.

\bibitem[{Rumyantsev} et~al.(2008){Rumyantsev}, {Antoniuk}, and
  {Pozanenko}]{crao2}
A.~{Rumyantsev}, K.~{Antoniuk}, and A.~{Pozanenko}.
\newblock {GRB\,080603B: optical observations in CrAO}.
\newblock \emph{GCN Circular 7974}, 2008.

\bibitem[{Rumyantsev} and {Pozanenko}(2008)]{crao}
V.~{Rumyantsev} and A.~{Pozanenko}.
\newblock {GRB\,080603B: optical observation}.
\newblock \emph{GCN Circular 7869}, 2008.

\bibitem[{Schlegel} et~al.(1998){Schlegel}, {Finkbeiner}, and {Davis}]{dust}
D.~{Schlegel}, D.~{Finkbeiner}, and M.~{Davis}.
\newblock {Maps of Dust Infrared Emission for Use in Estimation of Reddening
  and Cosmic Microwave Background Radiation Foregrounds}.
\newblock \emph{\apj}, 500:\penalty0 525, June 1998.
\newblock \doi{10.1086/305772}.

\bibitem[{Xin} et~al.(2008){Xin}, {Feng}, {Zhai}, {Qiu}, {Wei}, {Hu}, {Deng},
  {Wang}, {Urata}, and {Zheng}]{xing}
L.~{Xin}, Q.~{Feng}, M.~{Zhai}, Y.~{Qiu}, J.~{Wei}, J.~{Hu}, J.~{Deng},
  J.~{Wang}, Y.~{Urata}, and W.~{Zheng}.
\newblock {GRB\,080603B: Xinglong EST observations}.
\newblock \emph{GCN Circular 7814}, 2008.

\bibitem[{Zhuchkov} et~al.(2008){Zhuchkov}, {Bikmaev}, {Sakhibullin},
  {Khamitov}, {Eker}, {Kiziloglu}, {Gogus}, {Burenin}, {Pavlinsky}, and
  {Sunyaev}]{turkish}
R.~{Zhuchkov}, I.~{Bikmaev}, N.~{Sakhibullin}, I.~{Khamitov}, Z.~{Eker},
  U.~{Kiziloglu}, E.~{Gogus}, R.~{Burenin}, M.~{Pavlinsky}, and R.~{Sunyaev}.
\newblock {GRB\,080603B: RTT150 optical observations, break in light curves}.
\newblock \emph{GCN Circular 7803}, 2008.

\end{thebibliography}
\bibliographystyle{abbrvnat}

\end{document}